\documentclass[aps,prc,twocolumn,showpacs,superscriptaddress,longbibliography,floatfix,10pt]{revtex4-1}
\usepackage{indentfirst}
\usepackage{bm}
\usepackage{dcolumn}
\usepackage{amssymb}
\usepackage{amsmath}
\usepackage{graphicx}
\usepackage{dcolumn}
\usepackage{xcolor}
\usepackage{fix-cm}
\usepackage{mathptmx}
\usepackage[T1]{fontenc}
\usepackage{epstopdf}
\usepackage[colorlinks,citecolor=blue,linkcolor=blue,anchorcolor=blue,filecolor=blue,urlcolor=blue]{hyperref}
\usepackage{multirow}
\usepackage[utf8]{inputenc}

\usepackage{CJK}

\makeatletter
\def\NAT@def@citea{\def\@citea{\NAT@separator}}
\makeatother

\begin{document}
\hyphenpenalty=5000
\tolerance=2000
\begin{CJK*}{UTF8}{}
\title{Examination of promising reactions with $^{241}$Am and $^{244}$Cm targets for the synthesis of new superheavy elements within the dinuclear system model with a dynamical potential energy surface}

\author{Xiang-Quan Deng~\CJKfamily{gbsn}(邓祥泉)}
\affiliation{CAS Key Laboratory of Theoretical Physics, Institute of Theoretical Physics,
      Chinese Academy of Sciences, Beijing 100190, China}
\affiliation{School of Physical Sciences, University of Chinese Academy of Sciences, Beijing 100049, China}

\author{Shan-Gui Zhou~\CJKfamily{gbsn}(周善贵)} \email[Corresponding author:~]{sgzhou@itp.ac.cn}
\affiliation{CAS Key Laboratory of Theoretical Physics, Institute of Theoretical Physics,
	Chinese Academy of Sciences, Beijing 100190, China}
\affiliation{School of Physical Sciences, University of Chinese Academy of Sciences, Beijing 100049, China}
\affiliation{School of Nuclear Science and Technology, University of Chinese Academy of Sciences, Beijing 100049, China}
\affiliation{Synergetic Innovation Center for Quantum Effects and Application, Hunan Normal University, Changsha, 410081, China}
\affiliation{Peng Huanwu Collaborative Center for Research and Education, Beihang University, Beijing 100191, China}

\date{\today}

\begin{abstract}
Two actinide isotopes, $^{241}$Am and $^{244}$Cm, produced and chemically purified by the HFIR/REDC complex at ORNL are candidates for target materials of heavy-ion fusion reaction experiments for the synthesis of new superheavy elements (SHEs) with $Z>118$. In the framework of the dinuclear system model with a dynamical potential energy surface (DNS-DyPES model), we systematically study the $^{48}$Ca-induced reactions that have been applied to synthesize SHEs with $Z=112$--118, as well as the hot-fusion reactions with $^{241}$Am and $^{244}$Cm as targets which are promising for synthesizing new SHEs with $Z=119$--122. Detailed results including the maximal evaporation residue cross section and the optimal incident energy for each reaction are presented and discussed.
\end{abstract}
\maketitle
\newpage
\clearpage
\cleardoublepage
\end{CJK*}

\section{INTRODUCTION}

Exploring the upper limits of nuclear charge and mass and synthesizing new superheavy elements (SHEs) have been at the forefront of nuclear physics research for decades \cite{Hofmann2000_RMP72-733,Zagrebaev2001_PRC65-014607,Gan2001_EPJA10-21,Gan2004_EPJA20-385,Morita2004_JPSJ73-2593,Oganessian2010_PRL104-142502,Hofmann2011_RA99-405,Zhang2012_CPL29-012502,Adamian2016_PPN47-387,Oganessian2017_PS92-023003,Yu2017_SciChinaPMA60-092011,Hagino2018_PRC98-014607,Nazarewicz2018_NatPhys14-537,Giuliani2019_RMP91-011001,Voinov2020_BRASP84-351}.
Both macroscopic-microscopic models and self-consistent microscopic models predict proton magic numbers beyond $Z=82$, e.g., 114, 120, and 126, and neutron magic number beyond $N=126$, i.e., 184; the nuclei with and around these magic numbers have relatively high stability with respect to spontaneous fission, forming an ``island of stability'' of superheavy nuclei (SHN) beyond the mainland in the chart of nuclides \cite{Myers1966_NP81-1,Wong1966_PL21-688,Sobiczewski1966_PL22-500,Strutinsky1966_YF3-614,Meldner1967_ArkivF36-593,Nilsson1969_NPA131-1,Mosel1969_ZPA222-261,Rutz1997_PRC56-238,Decharge2003_NPA716-55,Zhang2005_NPA753-106,Sobiczewski2007_PPNP58-292,Zhang2012_PRC85-014325,Koura2013_JPSJ82-014201,Mo2014_PRC90-024320,Li2014_PLB732-169}. In addition, a shoal of SHN around $Z=108$ and $N=162$, in between the mainland and the elusive island of stability, has been predicted and studied extensively \cite{Moller1974_NPA229-292,Cwiok1983_NPA410-254,Patyk1989_NPA502-591,Patyk1991_NPA533-132,Smolanczuk1995_PRC52-1871,Meng2020_SciChinaPMA63-212011,Wang2022_CPC46-024107}. 
Meanwhile, much efforts have been made to study the synthesis mechanism of SHN both theoretically and experimentally.

Currently, as is known to us, the fusion-evaporation reaction of heavy-ions is a feasible approach to synthesize SHN (new SHEs in some cases) in the laboratories. By means of cold-fusion reactions, SHEs with $Z=107$--113 have been successfully synthesized \cite{Hofmann2000_RMP72-733,Morita2004_JPSJ73-2593,Hofmann2011_RA99-405}, while those with $Z=112$--118 have been produced via hot-fusion reactions \cite{Oganessian2017_PS92-023003}. A number of theoretical models have been developed to describe heavy-ion fusion-evaporation reactions for the synthesis of SHN. Based on different models, such as the dinuclear system (DNS) models \cite{Adamian2003_PRC68-034601,Feng2007_PRC76-044606,Adamian2009_EPJA41-235,Feng2009_NPA816-33,Nasirov2009_PRC79-024606,Gan2011_SciChinaPMA54-61,Nasirov2011_PRC84-044612,Kuzmina2012_PRC85-014319,Wang2012_PRC85-041601R,Adamian2014_PPN45-848,Bao2015_PRC91-011603R,Bao2015_PRC92-014601,Bao2015_PRC92-034612,Bao2016_JPG43-125105,Li2018_PRC98-014618,Adamian2020_PRC101-034301,Yang2020_NPR37-151,Adamian2021_EPJA57-89,Niu2021_NST32-103,Kayumov2022_PRC105-014618,Li2022_PRC105-054606}, the multidimensional Langevin-type dynamical equations \cite{Zagrebaev2008_PRC78-034610,Zagrebaev2012_PRC85-014608}, the fusion-by-diffusion models \cite{Liu2009_PRC80-054608,Siwek-Wilczynska2010_IJMPE19-500,Liu2011_PRC83-044613,Liu2011_PRC84-031602R,Siwek-Wilczynska2012_PRC86-014611,Liu2013_PRC87-034616,Liu2014_PRC89-024604,Siwek-Wilczynska2019_PRC99-054603}, the two-step model \cite{Liu2016_EPJA52-35}, and several empirical approaches \cite{Loveland2007_PRC76-014612,Wang2011_PRC84-061601R,Zhang2013_NPA909-36,Zhu2014_PRC89-024615,Ghahramany2016_EPJA52-287,Santhosh2016_PRC94-024623,Santhosh2017_PRC96-034610,Lv2021_PRC103-064616,Naderi2021_ChinPC45-094105,Wang2021_PRC103-034605}, many heavy-ion reactions have been investigated for the synthesis of new SHEs with $Z>118$. One of the main challenges in applying hot-fusion reactions to synthesize SHEs beyond Og is that the evaporation residue (ER) cross sections are extremely small---it has been shown that the upper limits are merely several tens of fb.

Among these theoretical models, our research interests lie in the DNS model. In this model, after they overcome the Coulomb barrier the two reacting nuclei touch each other and form a DNS. The two nuclei in the DNS keep their individualities and nucleon(s) may transfer from one nucleus to the other. The concept of DNS was firstly proposed to study the mechanism in deep inelastic heavy-ion collisions \cite{Volkov1978_PR44-93}. This concept was later adopted to deal with the competition between complete fusion and quasifission during the fusion process \cite{Antonenko1993_PLB319-425,Antonenko1995_PRC51-2635}. By considering the evolution of dynamical deformations of the two nuclei in the DNS, the dinuclear system model with a dynamical potential energy surface (DNS-DyPES model) was developed and used to describe the fusion dynamics and synthesis of SHN \cite{Wang2012_PRC85-041601R}. Some hot-fusion reactions for synthesizing SHEs with $Z=118$--120 have been studied with this model \cite{Wang2012_PRC85-041601R,Wang2014_PRC89-037601}.

In the cold-fusion reactions, $^{208}$Pb and $^{209}$Bi are used as targets, while in the hot-fusion reactions, $^{48}$Ca projectiles are adopted. In an attempt to synthesize new SHEs with $Z=119$ and 120 at GSI, Darmstadt \cite{Khuyagbaatar2020_PRC102-064602}, the $^{50}$Ti isotope was chosen as the projectile material. In that experiment, neither of these targeted SHEs was observed. Clearly, other than the currently tested reaction systems, more feasible combinations of projectile and target nuclei are in demand for the synthesis of new SHEs.
Besides $^{50}$Ti, the heaviest stable isotopes of several elements, namely $^{54}$Cr, $^{55}$Mn, $^{58}$Fe, and $^{59}$Co, can be alternatives for the $^{48}$Ca projectiles. In the meanwhile, there are already inventories of gram quantities of $^{241}$Am and $^{244}$Cm at ORNL which were produced and purified by the HFIR/REDC complex \cite{Roberto2015_NPA944-99,Robinson2020_RA108-737}. These two isotopes are promising candidates for targets as substitutes for the tested $^{243}$Am and $^{245,248}$Cm. The above-mentioned candidates for projectile and target nuclei provide a series of promising reactions for synthesizing new SHEs with $Z=119$--122. In this work, we study these reactions within the DNS-DyPES model and present detailed results.

The paper is organized as follows. The theoretical framework of the DNS-DyPES model is briefly formulated in Sec.~\ref{sec2}. In Sec.~\ref{sec3}, we present systematic calculation results and discussion of some $^{48}$Ca-induced hot-fusion reactions for the synthesis of SHEs with $Z=112$--118 and reactions with $^{241}$Am and $^{244}$Cm as targets for $Z=119$--122. We make a summary of this work in Sec.~\ref{sec4}.

\section{THEORETICAL FRAMEWORK}\label{sec2}

The synthesis process of a SHN by a heavy-ion fusion reaction can generally be divided into three successive stages \cite{Adamian1998_NPA633-409}. In the first stage a capture process occurs, namely, the projectile and target nuclei overcome the Coulomb barrier between them, leading to the formation of a composite system. In the second stage, the system undergoes fusion towards a compound nucleus (CN), which competes against quasifission.
In the third stage, the excited CN cools down through neutron emission and survives against fission. Within such a theoretical framework, we calculate the ER cross section as sum over partial waves,
\begin{equation}
\sigma_{\text{ER,~}xn}(E_{\text{c.m.}})=\sum_{J}\sigma_{\text{cap}}(E_{\text{c.m.}},J)
P_{\text{CN}}(E_{\text{c.m.}},J)W_{\text{sur,~}xn}(E_{\text{c.m.}},J),
\label{eq:eq3}
\end{equation}
where $E_\text{c.m.}$ is the incident energy in the center-of-mass frame and $J$ is the relative angular momentum between the projectile and target nuclei. In this work, we calculate the capture cross section $\sigma_{\text{cap}}$ with an empirical coupled-channel approach. The fusion probability $P_{\text{CN}}$ is determined within the DNS-DyPES model \cite{Wang2012_PRC85-041601R}. The survival probability $W_{\text{sur,~}x\text{n}}$ is calculated with a statistical model.

The capture process can be treated as a problem of penetration through the Coulomb barrier between the nuclei. The barrier splits into a set of discrete barriers when the relative motion is strongly coupled with the inelastic excitation channels \cite{Dasgupta1998_ARNPS48-401}. Such effect can be treated with empirical coupled-channel models \cite{Zagrebaev2001_PRC65-014607,Feng2006_NPA771-50,Wang2017_ADNDT114-281} in which the dynamical deformations of the interacting nuclei are taken into account. The capture cross section for a partial wave $J$ reads
\begin{equation}
\sigma_{\text{cap}}(E_{\text{c.m.}},J)=\frac{\pi\hbar^2}{2\mu E_{\text{c.m.}}}(2J+1)
T(E_{\text{c.m.}},J),
\end{equation}
where the transmission probability is formulated as
\begin{equation}
T(E_{\text{c.m.}},J)=\int dBf(B)\left\{1+\text{exp}\left[\frac{2\pi}{\hbar\omega_\text{B}(J)}(B_{\text{eff}}-E_{\text{c.m.}})\right]\right\}^{-1},
\end{equation}
\begin{equation}
B_{\text{eff}}=B+\frac{\hbar^2}{2\mu R_\text{B}^2}J(J+1).
\end{equation}
$\mu$ is the reduced mass of the nuclei, $B$ and $R_\text{B}$ are the height and the position of the barrier, $B_{\text{eff}}$ is the height of effective barrier, $\hbar\omega_\text{B}$ is the width of the barrier under the parabolic approximation. We take an asymmetric Gaussian function \cite{Zagrebaev2001_PRC65-014607}
\begin{equation}
f(B)=\frac{1}{N}\times
\left\{
\begin{matrix}
\text{exp}\left[-\left(\frac{B-B_\text{m}}{\Delta_{1}}\right)^2\right],~B<B_\text{m}\\
\text{exp}\left[-\left(\frac{B-B_\text{m}}{\Delta_{2}}\right)^2\right],~B>B_\text{m}\\
\end{matrix}
\right.,
\end{equation}
as the form of the barrier distribution, with $B_\text{m}=(B_{0}+B_\text{s})/2$, $\Delta_{2}=(B_{0}-B_\text{s})/2$, and $\Delta_{1}=\Delta_{2}-2$ MeV. $B_{0}$ is the height of the barrier between two colliding nuclei without dynamical deformations. When the dynamical deformation is considered, a two-dimensional potential energy surface can be calculated, and the height of the saddle point is $B_\text{s}$. $B_\text{m}$ is the central value of the barrier distribution, and the function $f(B)$ is normalized with the coefficient $N$. The nuclear and Coulomb parts of the potential are calculated with the Wong formula \cite{Wong1973_PRL31-766}. The capture cross section of the reaction can be calculated as
\begin{eqnarray}
\sigma_{\text{capture}}(E_{\text{c.m.}})=\sum_{J}\sigma_{\text{cap}}(E_{\text{c.m.}},J).
\end{eqnarray}

The fusion probability is calculated with the DNS-DyPES model \cite{Wang2012_PRC85-041601R}. Within the DNS framework, once the projectile is captured by the target, an initial DNS forms. A series of nucleon transfer occurring successively in such touching configurations may finally lead to the complete fusion of the system. This process is modeled by a diffusion of the DNS in the mass asymmetry degree of freedom, defined as $\eta=(A_1-A_2)/(A_1+A_2)$ ($A_1$ and $A_2$ are mass numbers of the nuclei composing the DNS). The DNS may also evolve along the relative distance of the nuclear centers of mass $R$, resulting in the decay of the system, namely quasifission. The evolution of the DNS dominated by the two competing mechanisms can be described by a master equation,
\begin{eqnarray}
\frac{dP(A_1,t)}{dt}=&&\sum_{A_{1}^{'}}W_{A_{1}A_{1}^{'}}(t)[d_{A_{1}}(t)P(A_{1}^{'},t)-d_{A_{1}^{'}}(t)P(A_{1},t)]\nonumber\\
&&-\Lambda_{A_1}^{\text{qf}}(t)P(A_{1},t).
\end{eqnarray}
Note that the total mass number of the DNS is constant, thus the mass number of one nucleus is enough to distinguish different DNSs. $P(A_1,t)$ is the probability distribution function of the DNS($A_1$, $A_2$) at time $t$. $W_{A_{1}A_{1}^{'}}(t)$ is the mean transition probability between the DNS($A_1$, $A_2$) and the DNS($A_1^{'}$, $A_2^{'}$). $d_{A_{1}}(t)$ is the microscopic dimension. $\Lambda_{A_1}^{\text{qf}}(t)$ is the quasifission rate along the $R$ degree of freedom. For a more detailed interpretation of the master equation and its numerical solution methods, the readers are referred to Refs.~\cite{Li2003_EPL64-750,Feng2006_NPA771-50}.

The dynamical deformations of the two colliding nuclei have a significant impact on the driving potential of the system during the fusion process. In the DNS-DyPES model, we assume that the dynamical deformation $\delta\beta(t)$ evolves in an overdamped form,
\begin{equation}
\delta\beta(t)=\delta\beta_{\text{max}}(1-e^{-t/\tau_{\text{def}}}).
\end{equation}
By minimizing the total intrinsic energy of the DNS, we determine the maximal deformation $\delta\beta_{\text{max}}$. The relaxation time for shape $\tau_{\text{def}}=40\times10^{-22} $ s \cite{Wolschin1979_PL88B-35}. As a consequence of the shape relaxation, the driving potential $V$ evolves in the same time scale. A formula is applied in numerical procedure to reduce computing cost,
\begin{equation}
V(t)=V(t=0)-\frac{\delta\beta(t)}{\delta\beta_{\text{max}}}[V(t=0)-V(t=\infty)],
\end{equation}
under the assumption that the potential energy varies linearly with the deformation.

The fusion probability for a partial wave $J$ can be calculated as the probability sum over the DNSs with $A_{1}$ no greater than that at the Businaro-Gallone (BG) point,
\begin{equation}
P_{\text{CN}}(E_{\text{c.m.}},J)=\sum\limits_{A_{1}=0}^{A_{1}=A_\text{BG}}P(E_{\text{c.m.}},J;A_{1},\tau_\text{int}),
\end{equation}
where $\tau_\text{int}$ is the interaction time of the nuclei in the DNS. Then, we can calculate the fusion cross section of the reaction as
\begin{eqnarray}
\sigma_{\text{fusion}}(E_{\text{c.m.}})=\sum_{J}\sigma_{\text{cap}}(E_{\text{c.m.}},J)
P_{\text{CN}}(E_{\text{c.m.}},J).
\end{eqnarray}
At a certain incident energy $E_\text{c.m.}$, the fusion probability averaging all partial waves is defined as
\begin{equation}
P_{\text{CN}}(E_{\text{c.m.}})=\frac{\sigma_{\text{fusion}}(E_{\text{c.m.}})}{\sigma_{\text{capture}}(E_{\text{c.m.}})}.
\end{equation}

The CN formed in a heavy-ion fusion reaction has a rather large excitation energy $E^{*}$. The excited nuclei may decay through fission or emission of light particles and $\gamma$ rays. The emission of light charged particles is strongly hindered by the Coulomb barrier, while the partial width for the emission of $\gamma$ rays is much smaller than that of the neutron at an excitation energy larger than neutron separation energy. Therefore, whether the hot CN may survive mainly depends on the competition between neutron evaporation and fission. Within a statistical model \cite{Zubov2002_PRC65-024308,Xia2011_SciChinaPMA54-s109}, the survival probability of a partial wave with $x$-neutron ($xn$) emission is formulated as
\begin{equation}
W_{\text{sur,~}xn}(E_{\text{c.m.}},J)=P_{\text{r.l.}}(E^{*},J,x)\prod_{i=1}^{x}\frac{\Gamma_\text{n}(E_{i}^{*},J)}{\Gamma_\text{n}(E_{i}^{*},J)+\Gamma_\text{f}(E_{i}^{*},J)}.
\end{equation}
The width of neutron emission $\Gamma_\text{n}$ is calculated with the evaporation model \cite{Feng2006_NPA771-50}. The width of fission $\Gamma_\text{f}$ is calculated with the Bohr-Wheeler formula \cite{Bohr1939_PR056-426}. The realization probability $P_{\text{r.l.}}$ is calculated by the method proposed in Ref.~\cite{Jackson1956_CanJP34-767}. The values of fission barrier and neutron separation energy are taken from Refs.~\cite{nrv,Sierk1986_PRC33-2039,Moller2016_ADNDT109--110-1}. We define the survival probability of the CN by averaging all partial waves,
\begin{equation}
W_{\text{sur,~}xn}(E_{\text{c.m.}})=\frac{\sigma_{\text{ER,~}xn}(E_{\text{c.m.}})}{\sigma_{\text{fusion}}(E_{\text{c.m.}})}.
\end{equation}

\section{RESULTS AND DISCUSSIONS}\label{sec3}

\begin{figure}[htbp]
\centering
\begin{tabular}{c}
\includegraphics[width=0.45\textwidth]{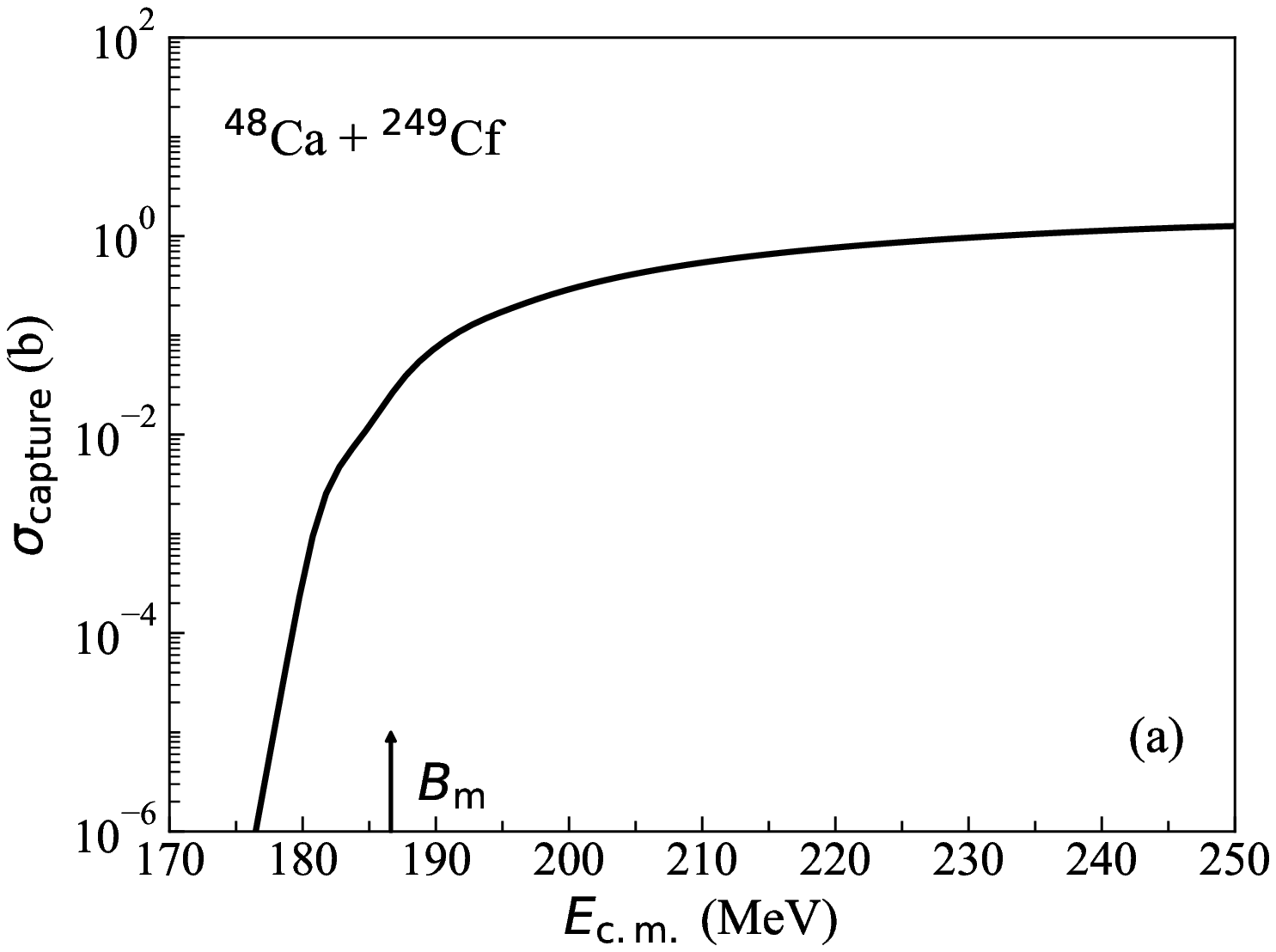}\\
\includegraphics[width=0.45\textwidth]{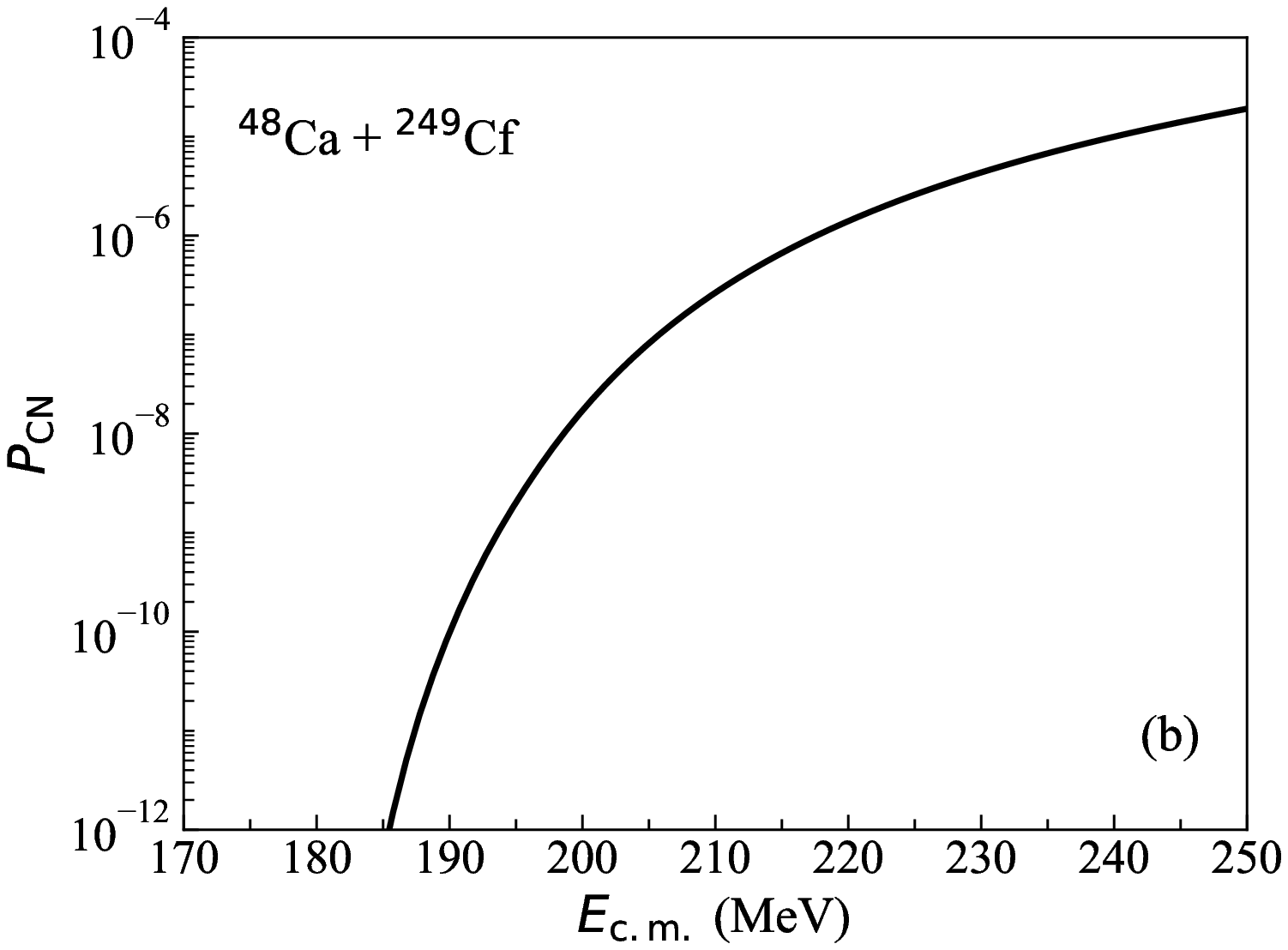}\\
\includegraphics[width=0.45\textwidth]{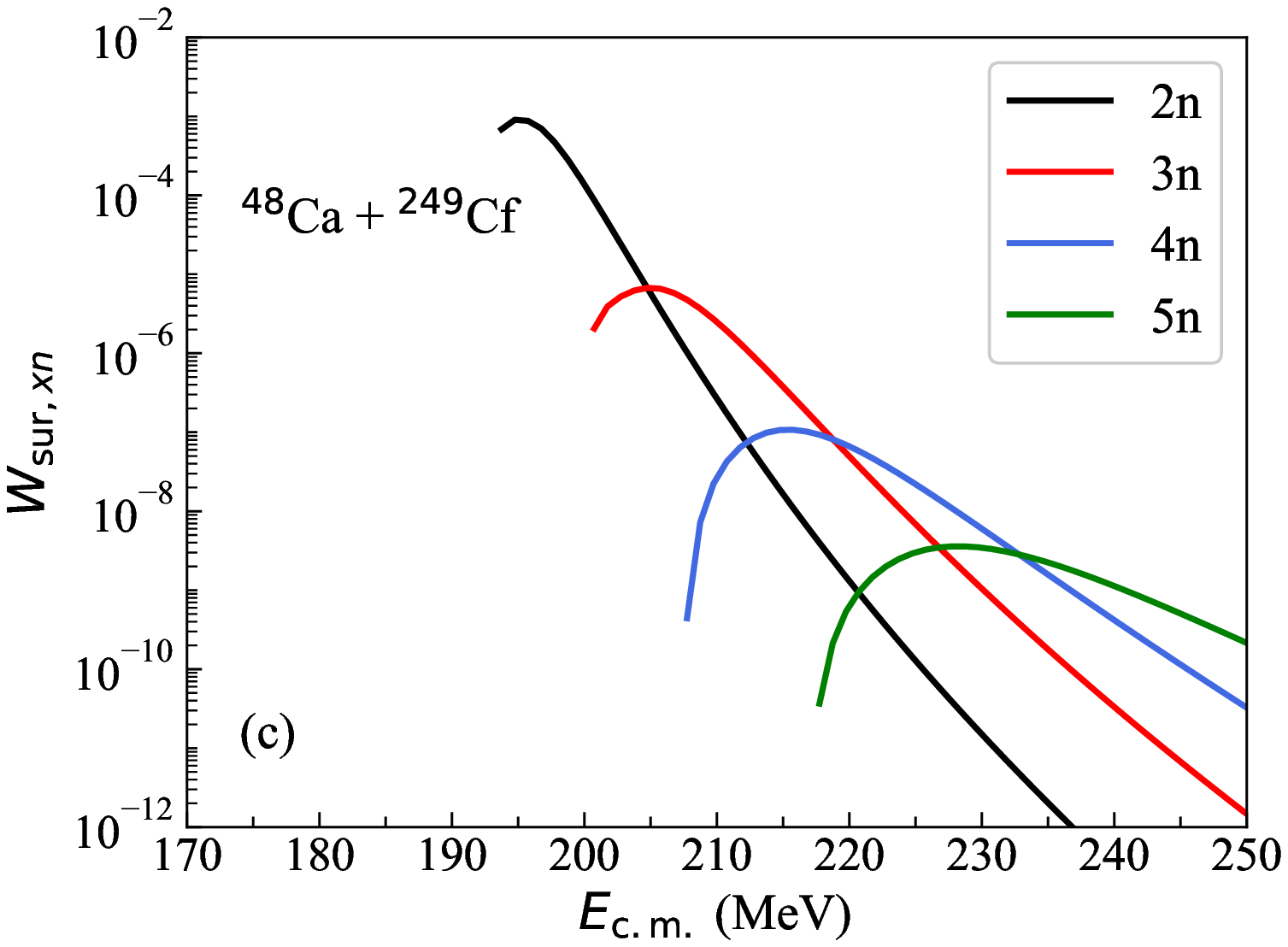}
\end{tabular}
\caption{(Color online) Capture cross section $\sigma_{\text{capture}}$ (a), fusion probability $P_{\text{CN}}$ (b), and survival probability $W_{\text{sur},xn}$ (c) as functions of the incident energy in the center-of-mass frame $E_{\text{c.m.}}$ for the reaction $^{48}$Ca + $^{249}$Cf.}\label{fig:1}
\end{figure}

\begin{figure}[htbp]
\centering
\begin{tabular}{c}
\includegraphics[width=0.45\textwidth]{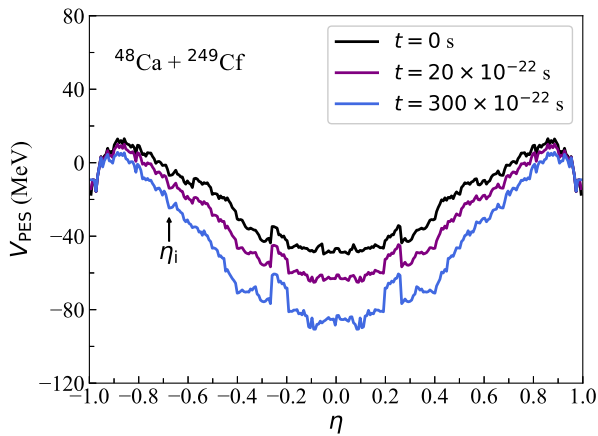}
\end{tabular}
\caption{(Color online) The evolution with time of the dynamical potential energy surface (DyPES) as functions of the mass asymmetry coordinate $\eta$ for the reaction $^{48}$Ca + $^{249}$Cf.}\label{fig:2}
\end{figure}

In this section, we systematically study the hot-fusion reactions which we mentioned in the introduction by using the DNS-DyPES model. Firstly, we investigate a typical reaction system, $^{48}$Ca + $^{249}$Cf, and illustrate the capture cross section, driving potential, fusion probability, and survival probability. Secondly, we study all the $^{48}$Ca-induced reactions that have been applied successfully to synthesize SHEs with $Z=112$--118 and compare the results with experimental data. Finally, we investigate reaction systems with promising candidates for the projectiles and targets in which new SHEs with $Z=119$--122 may be synthesized.

\subsection{A Typical Hot-fusion Reaction $^\mathbf{48}$Ca + $^\mathbf{249}$Cf}

We calculate the capture cross section, driving potential, fusion probability, and survival probability for the hot-fusion reaction $^{48}$Ca + $^{249}$Cf and show the results in Fig.~\ref{fig:1} and Fig.~\ref{fig:2}. At energies much lower than the central value of the barrier distribution $B_\mathrm{m}$, the reacting system mainly tunnels through the Coulomb barrier with pretty low tunneling probabilities, leading to very small capture cross sections, see Fig.~\ref{fig:1}(a). As the incident energy $E_\mathrm{c.m.}$ increases, the capture cross section dramatically increases around $B_\mathrm{m}$ and then saturates gradually.

In the DNS-DyPES model, the dynamical deformations of the projectile and the target are considered. When the dynamical deformations develop over interacting time of the two nuclei, the driving potential decreases as seen in Fig.~\ref{fig:2}, and the local excitation energies of DNSs increase.
As shown in Fig.~\ref{fig:1}(b), the DNS is more likely to fuse as the incident energy increases. This is because with a larger incident energy, the DNS has a greater possibility of getting over the inner fusion barrier along $\eta=(A_1-A_2)/(A_1+A_2)$ when diffusing. We can see that the complete fusion in such a system is quite rare to happen---the fusion probability is only $\approx10^{-5}$ when the incident energy is around 250 MeV.

As mentioned in the theoretical framework, the competition between neutron evaporation and fission determines the survival of the excited CN. Only when the excitation energy is larger than the one neutron separation energy may the CN cool down by emitting a neutron. Such thresholds in energy result in peaks in the survival probability, see Fig.~\ref{fig:1}(c).

\begin{figure*}[htbp]
\begin{tabular}{ccc}
\includegraphics[width=0.333\textwidth]{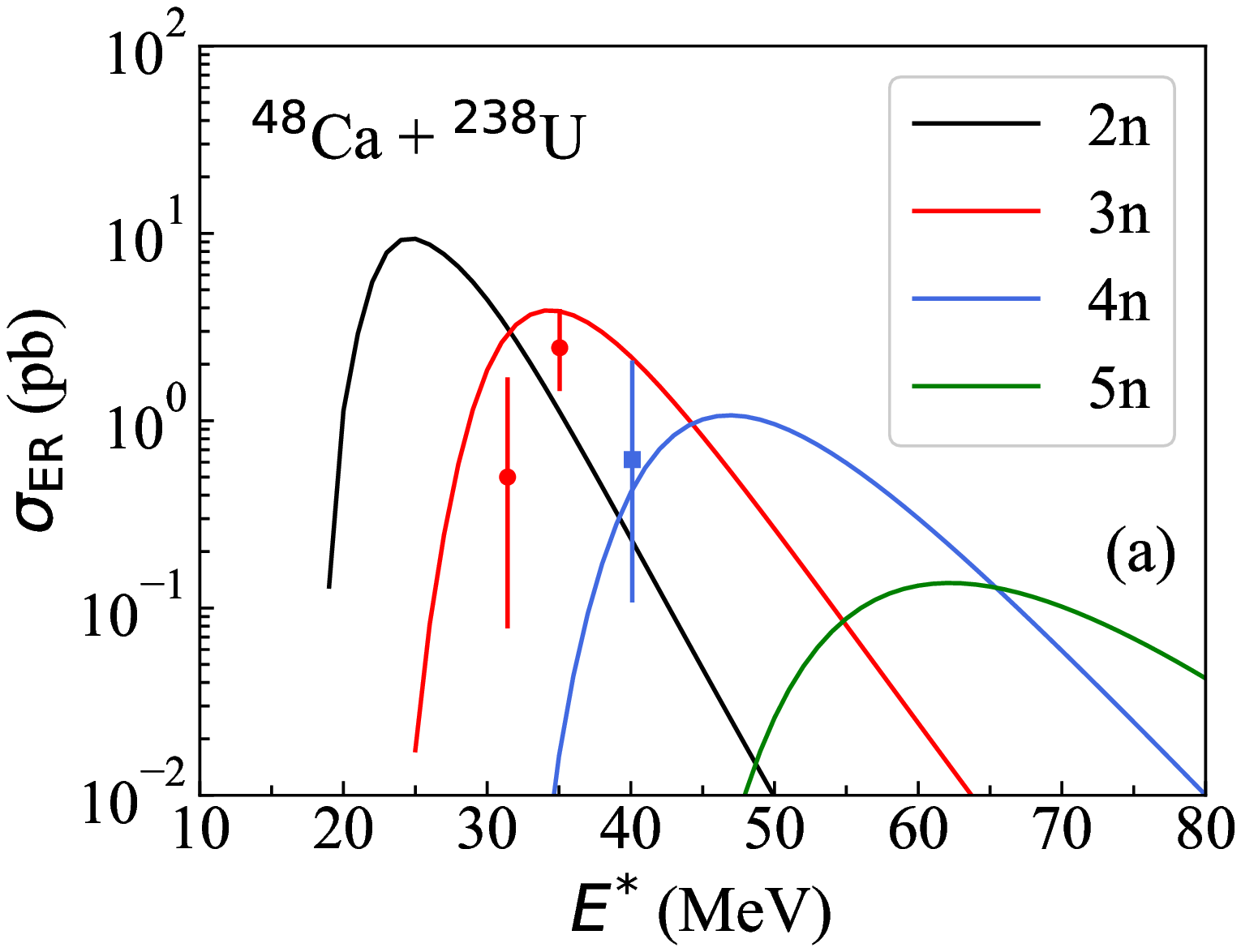}\includegraphics[width=0.333\textwidth]{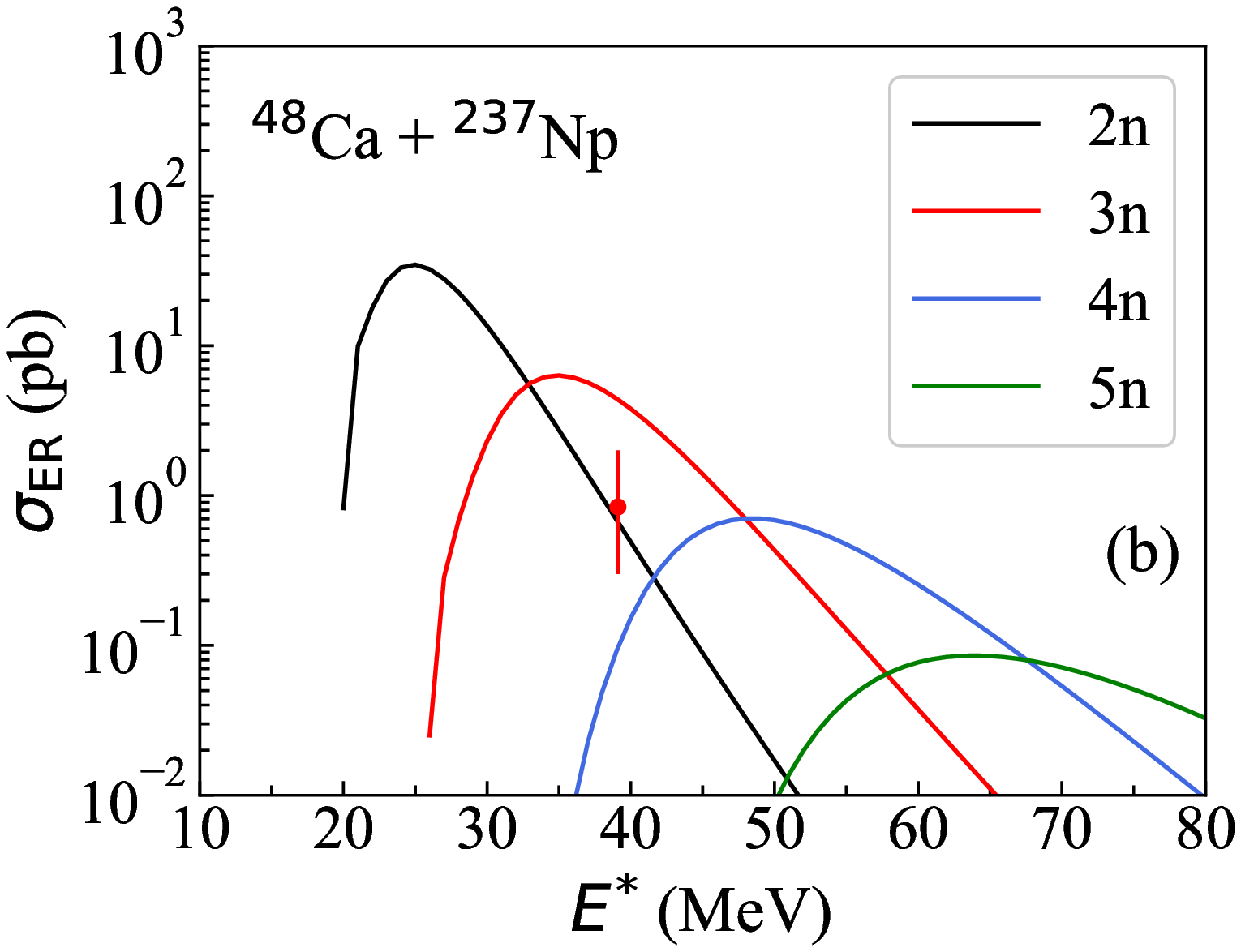}\includegraphics[width=0.333\textwidth]{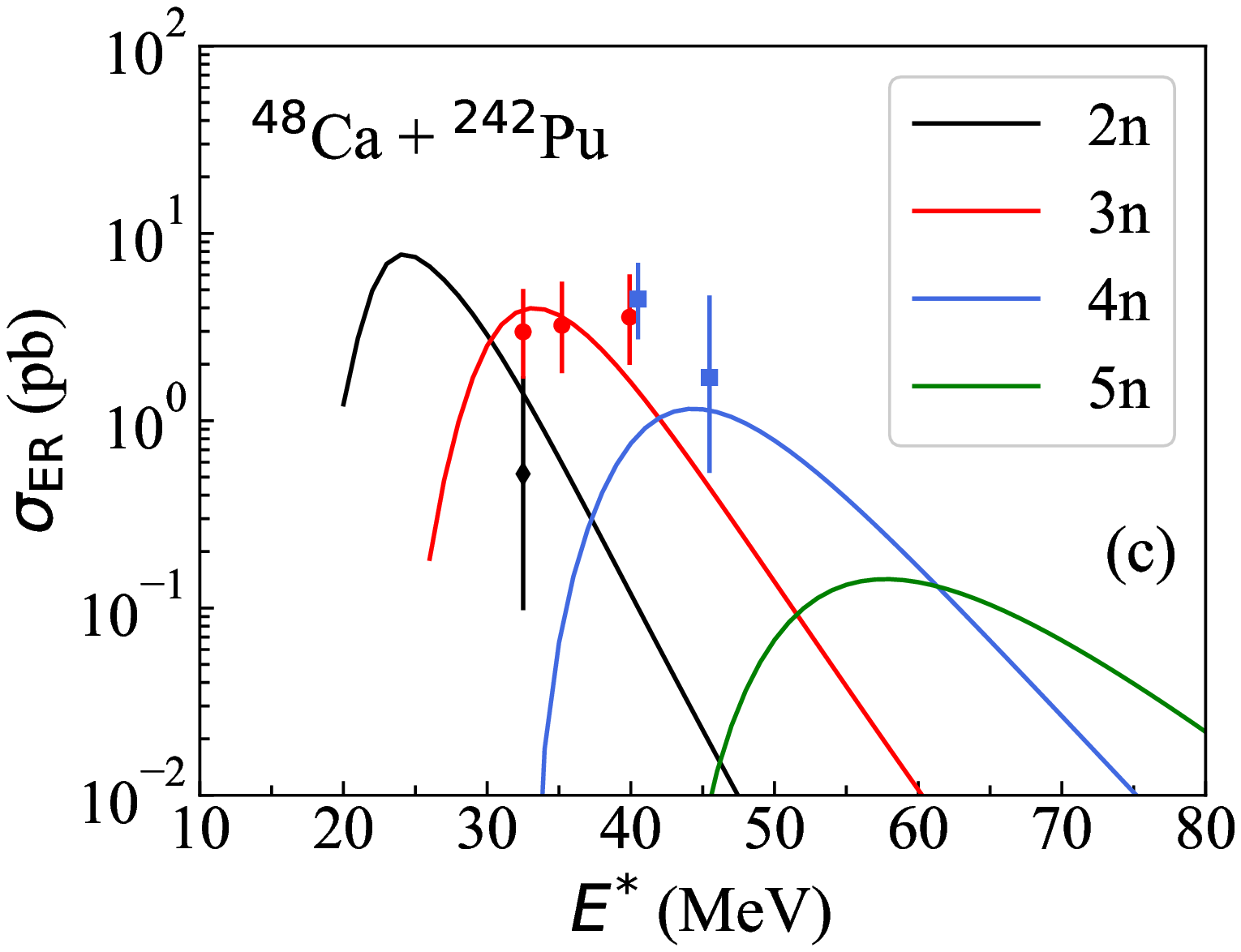}\\
\includegraphics[width=0.333\textwidth]{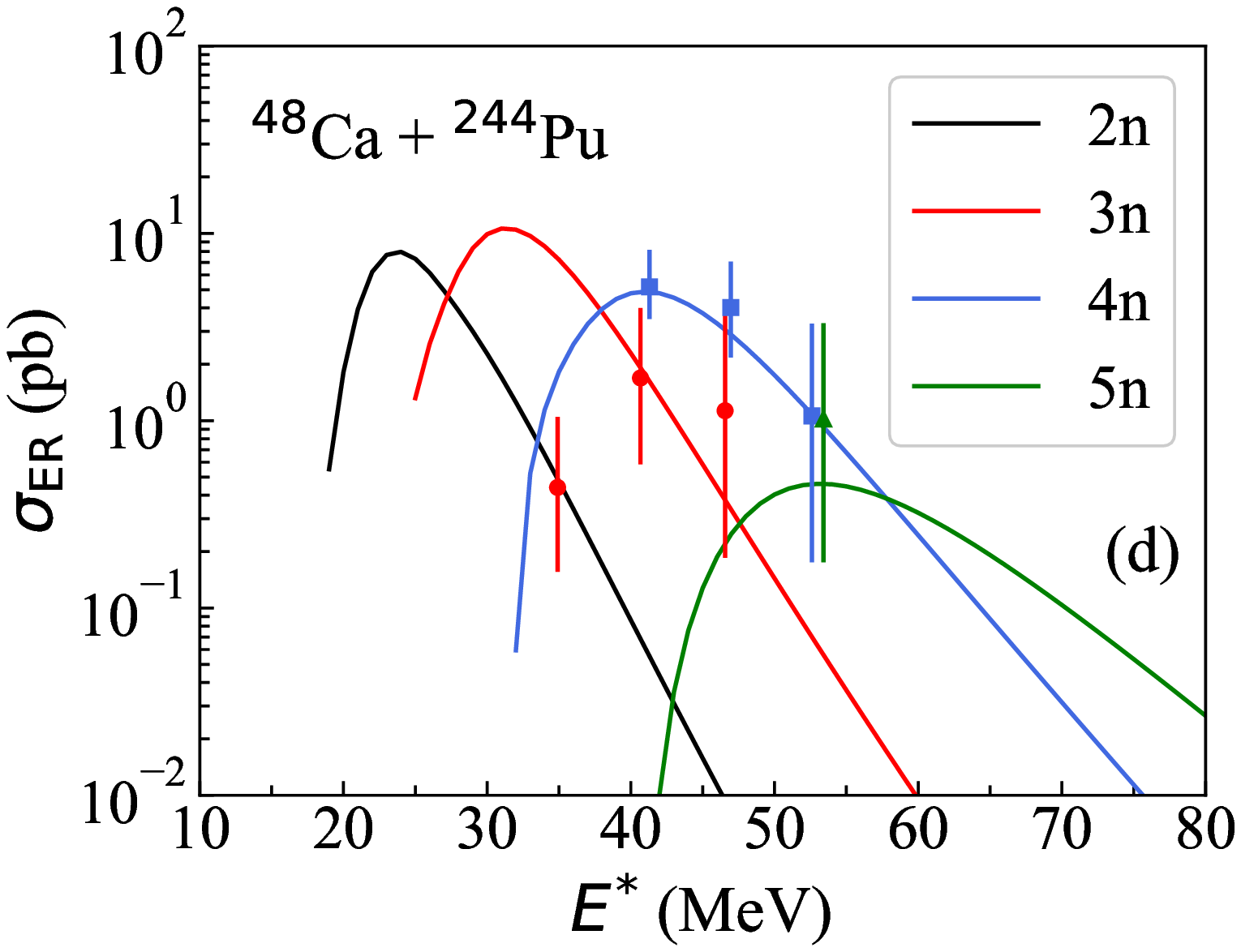}\includegraphics[width=0.333\textwidth]{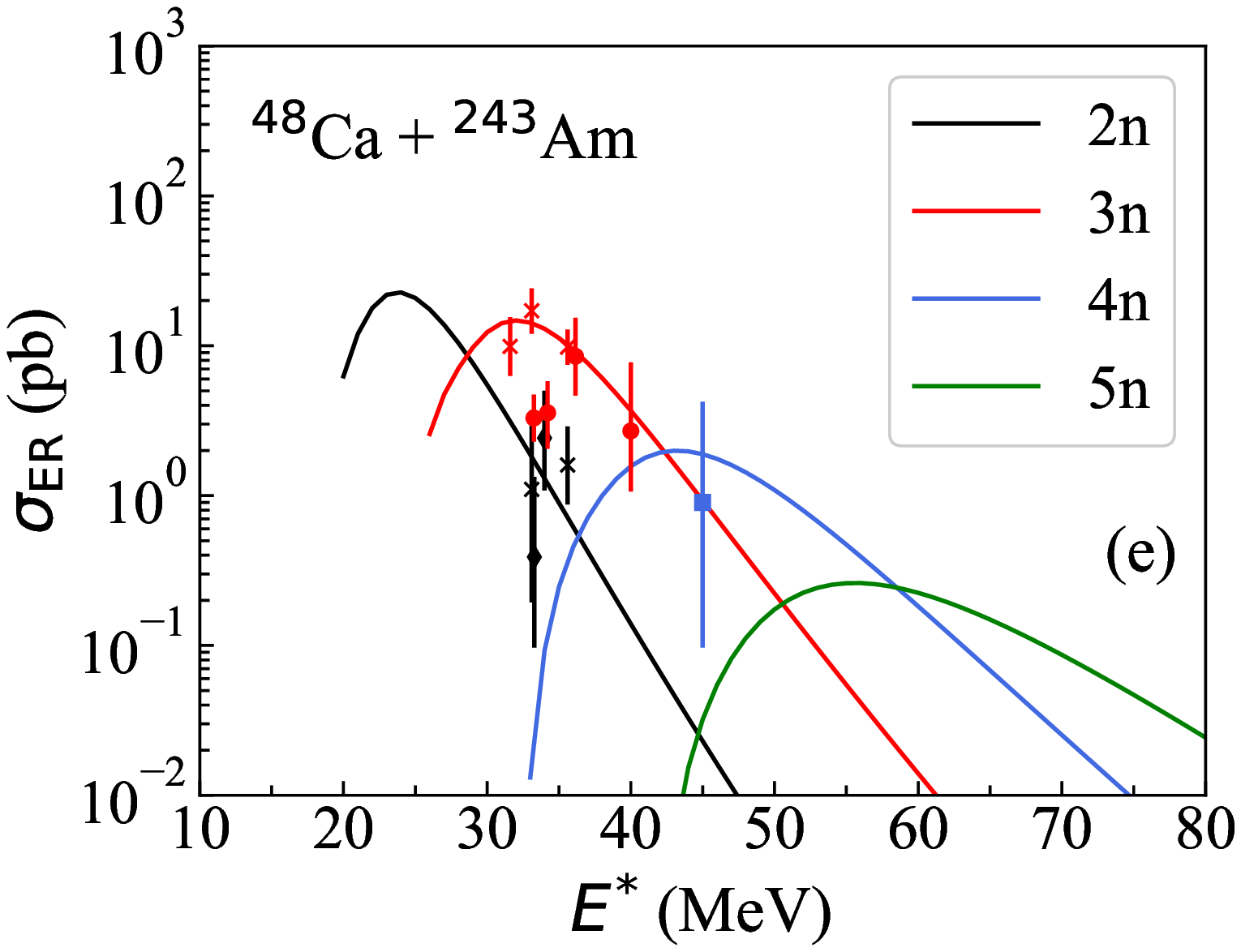}\includegraphics[width=0.333\textwidth]{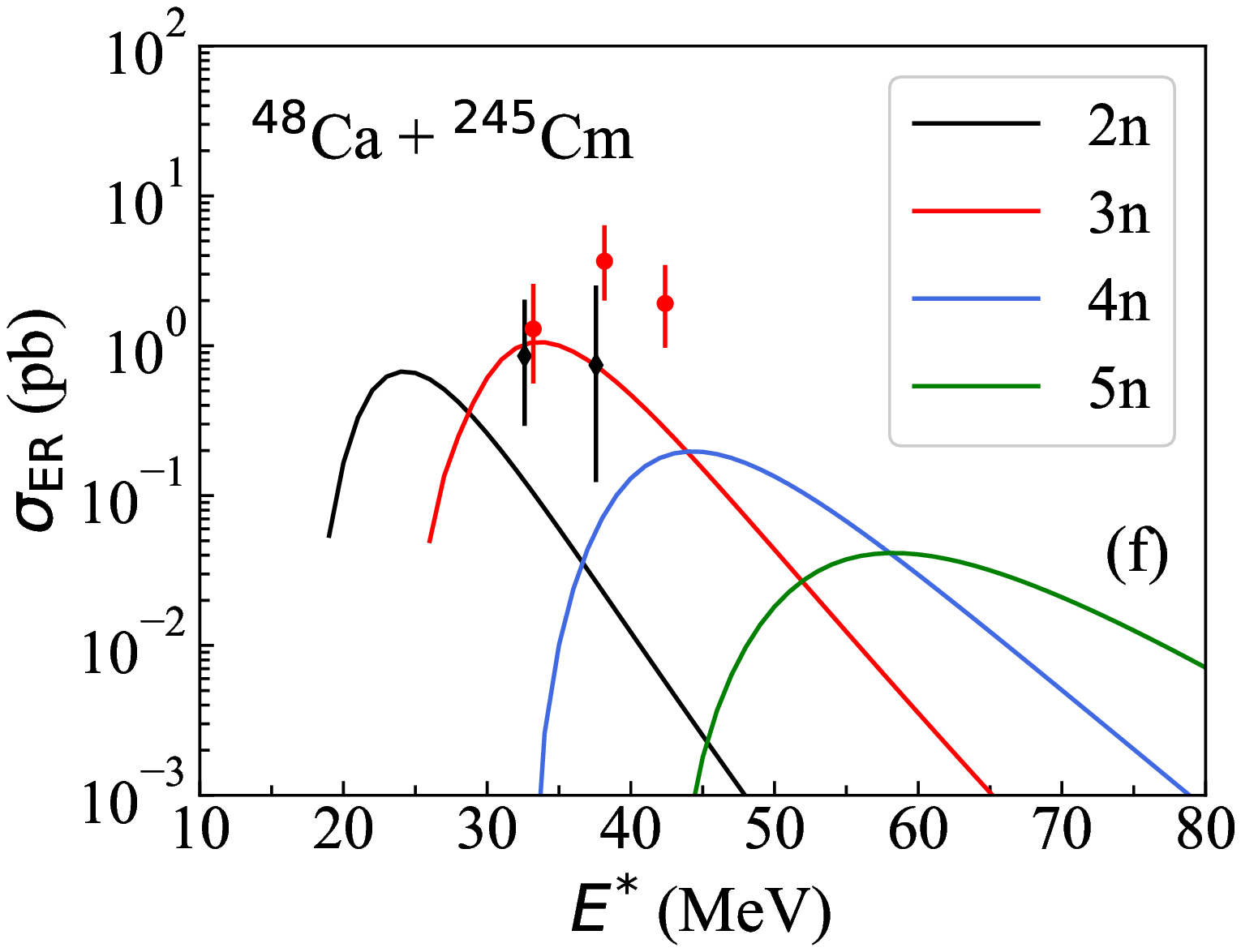}\\
\includegraphics[width=0.333\textwidth]{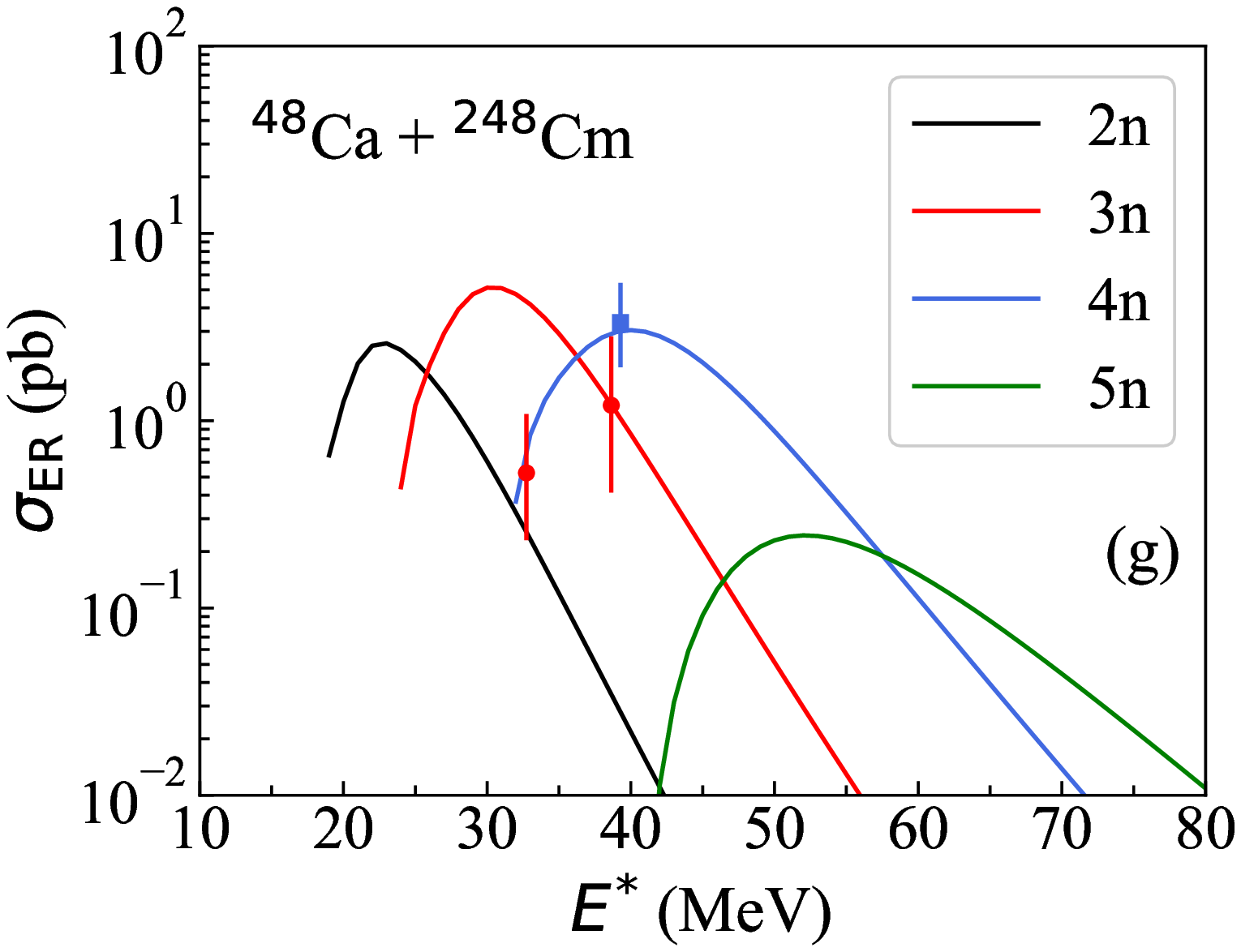}\includegraphics[width=0.333\textwidth]{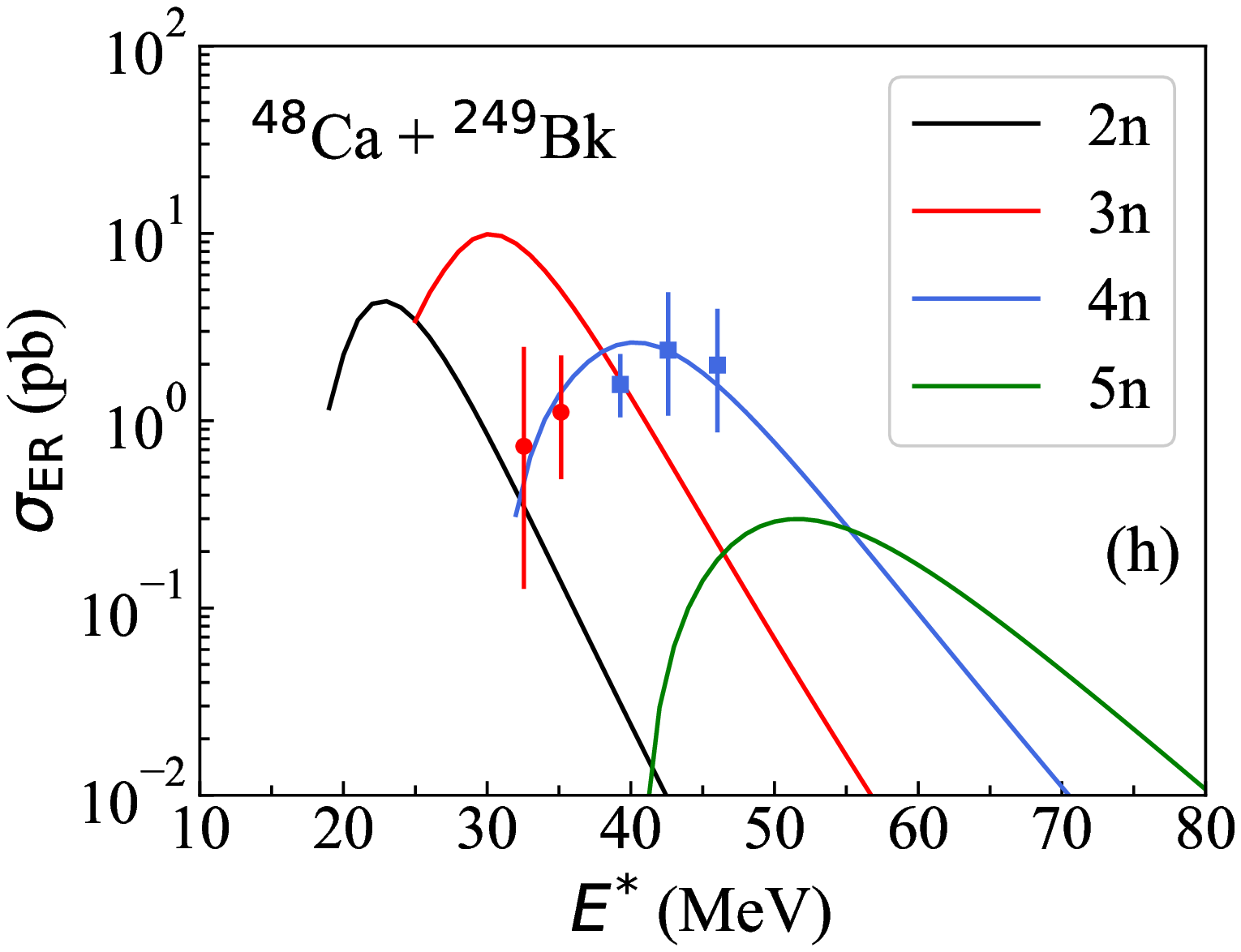}\includegraphics[width=0.333\textwidth]{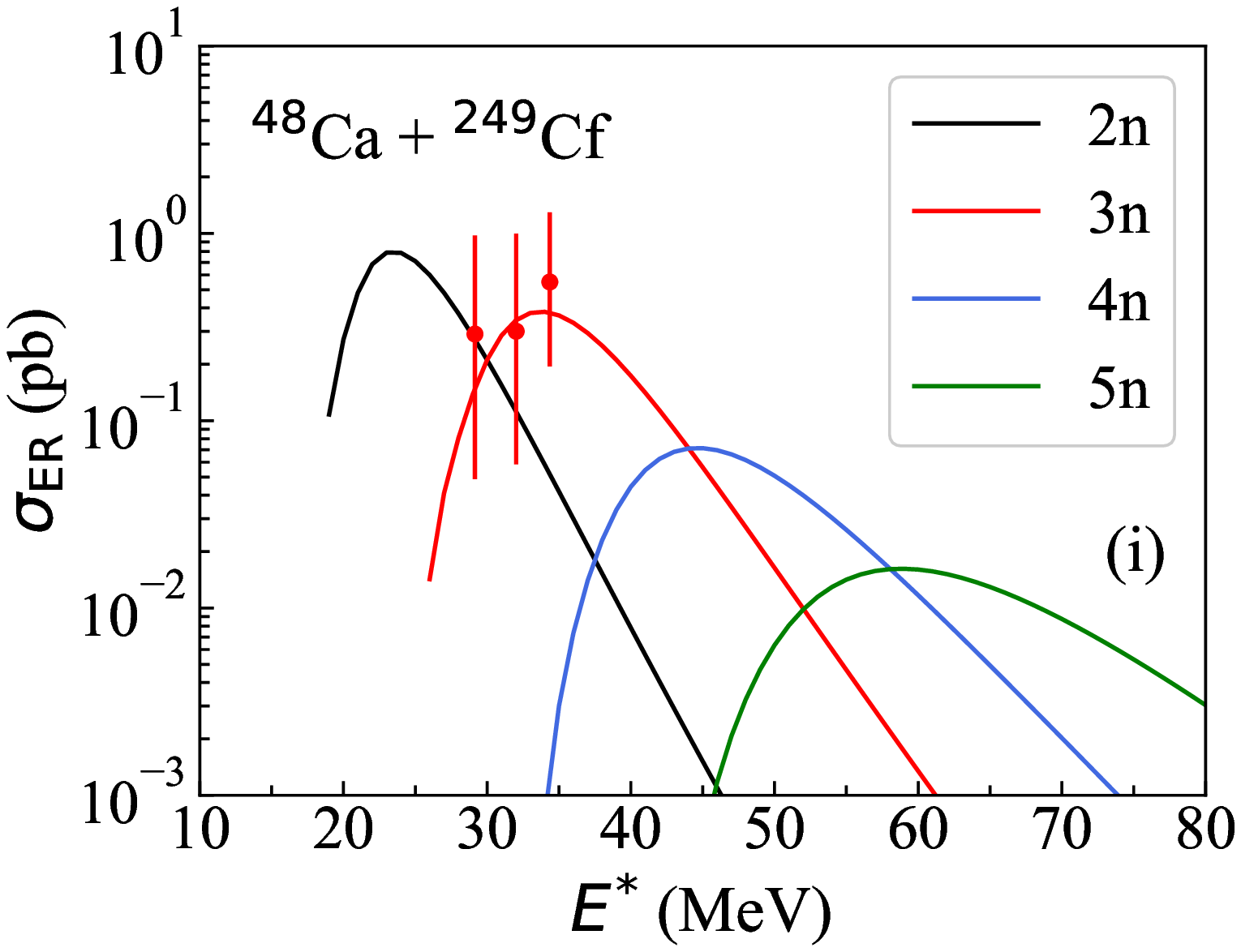}
\end{tabular}
\caption{(Color online) Calculated excitation functions of the $^{48}$Ca-induced reactions with target nuclei $^{238}$U, $^{237}$Np, $^{242}$Pu, $^{244}$Pu, $^{243}$Am, $^{245}$Cm, $^{248}$Cm, $^{249}$Bk, and $^{249}$Cf producing SHEs with $Z=112$--118, compared with experimental values of ER cross sections \cite{Oganessian2004_PRC69-054607,Oganessian2004_PRC70-064609,Oganessian2006_PRC74-044602,Oganessian2007_PRC76-011601R,Oganessian2013_PRC87-014302,Oganessian2013_PRC87-054621,Oganessian2015_RPP78-036301,Oganessian2022_PRC106-L031301}. Data taken from Refs.~\cite{Oganessian2004_PRC69-054607,Oganessian2004_PRC70-064609,Oganessian2006_PRC74-044602,Oganessian2007_PRC76-011601R,Oganessian2013_PRC87-014302,Oganessian2013_PRC87-054621,Oganessian2015_RPP78-036301} are shown by black diamonds, red circles, blues squares and green triangles for $2n$, $3n$, $4n$, and $5n$ channels, respectively. For $2n$ and $3n$ channels in the reaction $^{48}$Ca + $^{243}$Am, data given by a recent experiment \cite{Oganessian2022_PRC106-L031301} are shown by black and red crosses in (e).}\label{fig:3}
\end{figure*}

\subsection{$^\mathbf{48}$Ca-induced Reactions for SHEs with \textbf{\textit{Z}} = 112--118}

With a typical hot-fusion reaction illustrated, we now present the results of a systematic calculation of the excitation functions for the $^{48}$Ca-induced reactions with target nuclei $^{238}$U, $^{237}$Np, $^{242}$Pu, $^{244}$Pu, $^{243}$Am, $^{245}$Cm, $^{248}$Cm, $^{249}$Bk, and $^{249}$Cf and compare them with the available experimental values \cite{Oganessian2004_PRC69-054607,Oganessian2004_PRC70-064609,Oganessian2006_PRC74-044602,Oganessian2007_PRC76-011601R,Oganessian2013_PRC87-014302,Oganessian2013_PRC87-054621,Oganessian2015_RPP78-036301,Oganessian2022_PRC106-L031301} in Fig.~\ref{fig:3}. These hot-fusion reactions have been applied successfully in experiments to synthesize SHEs with $Z=112$--118 and are quite appropriate for the examination of theoretical calculations.

For $^{48}$Ca + $^{238}$U, the calculated ER cross sections of both $3n$ and $4n$ channels can reproduce the data well. There is only one experimental value for $^{48}$Ca + $^{237}$Np in the $3n$ channel, the calculation result is almost one order of magnitude larger than that value. With the $^{242}$Pu targets, the calculation results almost completely coincide with the data of $2n$, $3n$ and $4n$ channels. Experimental $\sigma_\mathrm{ER}$ values are available for $3n$, $4n$, and $5n$ channels of the reaction $^{48}$Ca + $^{244}$Pu. The calculated ER cross sections of the $3n$ channel reproduce two of the experimental values but there is a deviation in the optimal incident energy; for $4n$ and $5n$ channels, the calculated excitation functions and the data are in good agreement. For $^{48}$Ca + $^{243}$Am, when compared with the data taken from Ref.~\cite{Oganessian2013_PRC87-014302}, the excitation function of the $2n$ channel meets one of the data points; the calculated ER cross sections of the $3n$ channel reproduce two of the experimental values but the optimal incident energy of the $3n$ channel is underestimated; in the $4n$ channel, the calculated $\sigma_\mathrm{ER}$ agrees with the datum within the error bar. Benefiting from a new experimental complex at the SHE Factory at JINR, five new experimental values for $^{48}$Ca + $^{243}$Am in the $2n$ and $3n$ channels are reported recently \cite{Oganessian2022_PRC106-L031301}, shown by the black and red crosses in Fig.~\ref{fig:3}(e). Clearly, the calculated excitation functions can well reproduce the new experimental results. For $^{48}$Ca + $^{245}$Cm, the excitation function of the $2n$ channel is lower than the data points. In the $3n$ channel there are three data points. Our calculation curve meets one of them, but is lower than the other two. For $^{48}$Ca + $^{248}$Cm and $^{48}$Ca + $^{249}$Bk, the results in the $4n$ channel are in good agreement with the data; but the calculations of the $3n$ channel tend to overestimate the ER cross sections. The calculated $\sigma_\mathrm{ER}$'s of the $3n$ channel for the reaction $^{48}$Ca + $^{249}$Cf reproduce the data well.

On the whole, the DNS-DyPES model can describe the hot-fusion process and give a satisfactory description on the excitation functions of the reactions. We must note that, many well-established theories have hitherto given similar results, despite they are based on different physical scenarios when dealing with such complicated reaction process
\cite{Feng2006_NPA771-50,Zagrebaev2007_JPG34-1,Shen2008_IJMPE17-66,Wang2011_PRC84-061601R,Siwek-Wilczynska2012_PRC86-014611,Wang2012_PRC85-041601R,Zhang2013_NPA909-36,Zhu2014_PRC89-024615,Zhu2014_PRC90-014612,Bao2015_PRC91-011603R,Bao2015_PRC92-034612,Zagrebaev2015_NPA944-257,Bao2016_JPG43-125105,Hong2017_PLB764-42,Wu2018_PRC97-064609,Li2018_PRC98-014618,Lv2021_PRC103-064616}.

\begin{table*}[htbp]
\caption{Calculation results of the channels with the maximal ER cross section for some $^{48}$Ca-induced hot-fusion reactions and some new ones with $^{241}$Am and $^{244}$Cm targets. We choose the heaviest stable isotopes of elements with $Z=24$--27 as alternatives for $^{48}$Ca projectiles. For each reaction, the reaction channel with the maximal ER cross section, the optimal incident energy in the center-of-mass frame $E_\mathrm{c.m.}$, the optimal excitation energy $E^{*}$, the maximal ER cross section $\sigma_\mathrm{ER}$, and the corresponding capture cross section $\sigma_\mathrm{capture}$, fusion probability $P_\mathrm{CN}$, and survival probability $W_{\text{sur,~}xn}$ are listed.}\label{tab:table1}
\begin{ruledtabular}
\begin{tabular}{ldddlll}
\multicolumn{1}{c}{\textrm{Reaction Channel}}&\multicolumn{1}{c}{$E_\mathrm{c.m.}$} \textrm{(MeV)}&\multicolumn{1}{c}{$E^{*}$  \textrm{(MeV)}}&\multicolumn{1}{c}{$\sigma_\mathrm{capture}$ \textrm{(b)}}&\multicolumn{1}{c}{$P_\mathrm{CN}$}&\multicolumn{1}{c}{$W_{\text{sur,~}xn}$}&\multicolumn{1}{c}{$\sigma_\mathrm{ER}$ \textrm{(pb)}}\\
\colrule
$^{48}$Ca + $^{238}$U $\rightarrow$ $^{284}$Cn + $2n$ & 184.11 & 25.0 & 0.13 & $7.79\times10^{-7}$ & $9.33\times10^{-5}$ & $9.35$\\
$^{48}$Ca + $^{237}$Np $\rightarrow$ $^{283}$Nh + $2n$ & 190.82 & 25.0 & 0.23 & $2.24\times10^{-6}$ & $6.84\times10^{-5}$ & $34.7$\\
$^{48}$Ca + $^{242}$Pu $\rightarrow$ $^{288}$Fl + $2n$ & 186.62 & 24.0 & 0.12 & $7.75\times10^{-8}$ & $8.57\times10^{-4}$ & $7.71$\\
$^{48}$Ca + $^{244}$Pu $\rightarrow$ $^{289}$Fl + $3n$ & 192.29 & 31.0 & 0.25 & $4.29\times10^{-7}$ & $9.80\times10^{-5}$ & $1.06\times10^{1}$\\
$^{48}$Ca + $^{241}$Am $\rightarrow$ $^{287}$Mc + $2n$ & 194.75 & 24.0 & 0.26 & $2.57\times10^{-7}$ & $3.74\times10^{-4}$ & $2.47\times10^{1}$\\
$^{48}$Ca + $^{243}$Am $\rightarrow$ $^{289}$Mc + $2n$ & 192.08 & 24.0 & 0.20 & $1.39\times10^{-7}$ & $8.28\times10^{-4}$ & $2.27\times10^{1}$\\
$^{48}$Ca + $^{244}$Cm $\rightarrow$ $^{290}$Lv + $2n$ & 196.92 & 25.0 & 0.29 & $1.02\times10^{-7}$ & $1.87\times10^{-4}$ & $5.55$\\
$^{48}$Ca + $^{245}$Cm $\rightarrow$ $^{290}$Lv + $3n$ & 202.94 & 34.0 & 0.45 & $4.01\times10^{-7}$ & $5.87\times10^{-6}$ & $1.05$\\
$^{48}$Ca + $^{248}$Cm $\rightarrow$ $^{293}$Lv + $3n$ & 197.23 & 30.0 & 0.31 & $7.28\times10^{-8}$ & $2.26\times10^{-4}$ & $5.13$\\
$^{48}$Ca + $^{249}$Bk $\rightarrow$ $^{294}$Ts + $3n$ & 200.73 & 30.0 & 0.36 & $9.46\times10^{-8}$ & $2.93\times10^{-4}$ & $9.89$\\
$^{48}$Ca + $^{249}$Cf $\rightarrow$ $^{295}$Og + $2n$ & 197.75 & 23.0 & 0.24 & $7.11\times10^{-9}$ & $4.72\times10^{-4}$ & $7.90\times10^{-1}$\\
$^{54}$Cr + $^{241}$Am $\rightarrow$ $^{293}$119 + $2n$ & 233.22 & 26.0 & 0.45 & $1.25\times10^{-9}$ & $5.73\times10^{-5}$ & $3.20\times10^{-2}$\\
$^{54}$Cr + $^{244}$Cm $\rightarrow$ $^{296}$120 + $2n$ & 237.11 & 27.0 & 0.51 & $5.15\times10^{-10}$ & $1.24\times10^{-5}$ & $3.28\times10^{-3}$\\
$^{55}$Mn + $^{241}$Am $\rightarrow$ $^{294}$120 + $2n$ & 242.67 & 28.0 & 0.45 & $4.09\times10^{-9}$ & $9.22\times10^{-6}$ & $1.71\times10^{-2}$\\
$^{55}$Mn + $^{244}$Cm $\rightarrow$ $^{297}$121 + $2n$ & 245.92 & 27.0 & 0.50 & $2.70\times10^{-9}$ & $6.73\times10^{-6}$ & $9.15\times10^{-3}$\\
$^{58}$Fe + $^{241}$Am $\rightarrow$ $^{297}$121 + $2n$ & 255.88 & 27.0 & 0.56 & $4.08\times10^{-10}$ & $7.09\times10^{-6}$ & $1.61\times10^{-3}$\\
$^{58}$Fe + $^{244}$Cm $\rightarrow$ $^{300}$122 + $2n$ & 258.80 & 27.0 & 0.60 & $2.38\times10^{-10}$ & $4.75\times10^{-6}$ & $6.73\times10^{-4}$\\
$^{59}$Co + $^{241}$Am $\rightarrow$ $^{298}$122 + $2n$ & 263.59 & 28.0 & 0.45 & $2.00\times10^{-9}$ & $1.62\times10^{-6}$ & $1.45\times10^{-3}$\\
\end{tabular}
\end{ruledtabular}
\end{table*}

\begin{table}[htbp]
\caption{Calculation results of the $3n$ channels for the reactions $^{48}$Ca + $^{241}$Am and $^{48}$Ca + $^{244}$Cm with nuclear properties taken from different mass tables. For either reaction, the maximal ER cross sections $\sigma_\mathrm{ER}$, the optimal excitation energies $E^{*}$, the references, and the mass tables are listed.}\label{tab:table2}
\begin{ruledtabular}
\begin{tabular}{lddll}
\multicolumn{1}{c}{\textrm{Reaction}}&\multicolumn{1}{c}{$\sigma_\mathrm{ER}$ \textrm{(pb)}}&\multicolumn{1}{c}{$E^{*}$  \textrm{(MeV)}}&\multicolumn{1}{c}{\textrm{Reference}}&\multicolumn{1}{c}{\textrm{Mass Table}}\\
\colrule
$^{48}$Ca + $^{241}$Am & 7.31 & 33.0 & This work & \cite{nrv,Sierk1986_PRC33-2039,Moller2016_ADNDT109--110-1}\\
                       & 3.60 & 33.2 & \cite{Adamian2014_PPN45-848} & \cite{Moller1988_ADNDT39-213}\\
                       & 1.91 & 34.6 & \cite{Adamian2014_PPN45-848} & \cite{Moller1995_ADNDT59-185}\\
                       & 2.41 & 33.3 & \cite{Adamian2014_PPN45-848} & \cite{Myers1996_NPA601-141}\\
                       & 0.97 & 31.3 & \cite{Adamian2016_PPN47-387} & \cite{Adamian2016_PPN47-387}\\
$^{48}$Ca + $^{244}$Cm & 1.49 & 34.0 & This work & \cite{nrv,Sierk1986_PRC33-2039,Moller2016_ADNDT109--110-1}\\
                       & 0.98 & 31.7 & \cite{Adamian2014_PPN45-848} & \cite{Moller1988_ADNDT39-213}\\
\end{tabular}
\end{ruledtabular}
\end{table}

\subsection{New Reactions with Promising Projectiles and Targets Leading to SHEs with \textbf{\textit{Z}} = 119--122}

The combination of projectile and target nuclei is of great importance to the experiments in synthesizing new superheavy nuclei (SHN) and the maximal ER cross sections in various reactions may differ by several orders of magnitude. Hence, it is necessary to study untested reactions that can be carried out practically in laboratory as many as possible. For projectiles, the heaviest stable isotopes of elements with $Z=24$--27 are available alternatives for $^{48}$Ca. For targets, the actinide isotopes $^{241}$Am and $^{244}$Cm mentioned in the introduction are promising candidates. We study reaction systems with these isotopes that may lead to the synthesis of SHEs with $Z=119$--122.

For these reactions as well as $^{48}$Ca-induced ones discussed in the previous subsection, the detailed results of the channels with the maximal ER cross section for the reactions are listed in Table~\ref{tab:table1}. By investigating all the $xn$ channels of the reaction systems leading to SHEs with $Z=119$--122, we conclude that for each system, the ER cross section reaches a maximum when the CN with an excitation energy of 26.0--28.0 MeV cools down by emitting two neutrons. For the synthesis of SHE with $Z=119$, the maximal $\sigma_\mathrm{ER}$ in the reaction $^{54}$Cr + $^{241}$Am is found to be 32.0 fb. In the reactions $^{55}$Mn + $^{241}$Am and $^{54}$Cr + $^{244}$Cm, SHN $^{294}$120 and $^{296}$120 may be synthesized, respectively, in $2n$ evaporation channels. With the $^{241}$Am target, the maximal $\sigma_\mathrm{ER}$ is 17.1 fb; with the $^{244}$Cm target, the maximal $\sigma_\mathrm{ER}$ has a smaller value of 3.28 fb, however, a more neutron-rich SHN may be produced in this reaction. The projectile-target combinations $^{55}$Mn + $^{244}$Cm and $^{59}$Co + $^{241}$Am are favorable among the listed paths for synthesizing SHEs with $Z=121$ and 122, with the maximal $\sigma_\mathrm{ER}$'s being 9.15 fb and 1.45 fb.

For the $3n$ channels of the reactions $^{48}$Ca + $^{241}$Am and $^{48}$Ca + $^{244}$Cm, we obtain that the maximal ER cross sections are 7.31 pb with $E^{*}=33.0$ MeV and 1.49 pb with $E^{*}=34.0$ MeV. Based on the dinuclear system concept, Adamian et al. also studied these two reaction channels \cite{Adamian2014_PPN45-848,Adamian2016_PPN47-387} with the parameters of nuclear properties taken from several different mass tables \cite{Moller1988_ADNDT39-213,Moller1995_ADNDT59-185,Myers1996_NPA601-141,Adamian2016_PPN47-387}, see Table~\ref{tab:table2}. They obtained for the reaction channel $^{48}$Ca + $^{241}$Am $\rightarrow$ $^{286}$Mc + $3n$, $\sigma_\mathrm{ER}=3.60$ pb, 1.91 pb, 2.41 pb, and 0.97 pb with $E^{*}=33.2$ MeV, 34.6 MeV, 33.3 MeV, and 31.3 MeV, using mass tables given in Refs.~\cite{Moller1988_ADNDT39-213}, \cite{Moller1995_ADNDT59-185}, \cite{Myers1996_NPA601-141}, and \cite{Adamian2016_PPN47-387}, respectively; for $^{48}$Ca + $^{244}$Cm $\rightarrow$ $^{289}$Lv + $3n$, $\sigma_\mathrm{ER}=0.98$ pb with $E^{*}=31.7$ MeV using mass table in Ref.~\cite{Moller1988_ADNDT39-213}. Clearly, nuclear properties taken from different mass tables, including mass, neutron separation energy, and microscopic corrections, may significantly influence the prediction of the maximal ER cross sections and optimal excitation energies. The above-used nuclear properties are from macroscopic-microscopic models and we are looking forward to nuclear properties obtained with self-consistent methods \cite{Lu2014_PRC89-014323,Zhou2016_PS91-063008,Zhao2017_PRC95-014320}.

\section{SUMMARY}\label{sec4}

In summary, we systematically study a series of hot-fusion reactions for the synthesis of SHEs within the framework of the DNS-DyPES model. With the reaction $^{48}$Ca + $^{249}$Cf, we show the calculation results for the capture cross section, driving potential, fusion probability, and survival probability for a typical hot-fusion reaction system. During the fusion process, the development of the dynamical deformations of the reacting nuclei has a considerable impact on the driving potential. We calculate the excitation functions of the $^{48}$Ca-induced reactions that have been successfully applied in experiments to synthesize SHEs with $Z=112$--118 and the calculation results are in agreement with the experiments. Especially for the reaction $^{48}$Ca + $^{243}$Am, the excitation functions reproduce well the latest results given by the new experimental complex at the SHE Factory at JINR.

With two actinide isotopes $^{241}$Am and $^{244}$Cm as alternatives for currently tested targets, a series of promising reactions for the synthesis of new SHEs with $Z=119$--122 are proposed. We investigate these reactions and present detailed calculation results. We conclude that the reaction systems $^{54}$Cr + $^{241}$Am, $^{55}$Mn + $^{241}$Am, $^{55}$Mn + $^{244}$Cm, and $^{59}$Co + $^{241}$Am are the appropriate choices among the reactions we study in this work for synthesizing new SHEs with $Z=119$--122. The maximal $\sigma_\mathrm{ER}$'s in $2n$ evaporation channels are 32.0 fb, 17.1 fb, 9.15 fb, and 1.45 fb, respectively.

\begin{acknowledgments}
		Helpful discussions with
		Bing Wang, Bing-Nan Lu, and Xiang-Xiang Sun
		are gratefully acknowledged.
		This work has been supported by
		the National Key R\&D Program of China (Grant No. 2018YFA0404402),
		the National Natural Science Foundation of China (Grants
		No. 11525524, No. 12070131001, No. 12047503, and No. 11961141004),
		the Key Research Program of Frontier Sciences of Chinese Academy of Sciences (Grant No. QYZDB-SSWSYS013),
the Strategic Priority Research Program of Chinese Academy of Sciences (Grants No. XDB34010000 and No. XDPB15),
the Inter-Governmental S\&T Cooperation Project between China and Croatia,
and the IAEA Coordinated Research Project ``F41033''. The results described in this paper are obtained on the High-performance Computing Cluster of ITP-CAS and the ScGrid of the Supercomputing Center, Computer Network Information Center of Chinese Academy of Sciences.
\end{acknowledgments}

\bibliographystyle{apsrev4-1}
\bibliography{DNS_newreactions}
\end{document}